\documentclass[prl,twocolumn,showpacs,amsmath,amssymb]{revtex4}

\def\gfivehat{\hat{\gamma}_5}
\def\gfive{\gamma_5}

\def\eps{\varepsilon}
\def\tr{{\rm Tr}}

\def\Tr{\,{\rm Tr}\:}

\def\Nfour{${\cal N}{=}4$}
\def\NfourSuYM{${\cal N}{=}4$ Super-Yang-Mills}
\def\NfourSYM{${\cal N}{=}4$ SYM}
\def\nc{N_{\rm c}}

\def\st{\begin{equation}}
\def\stp{\end{equation}}
\def\bg{\begin{eqnarray}}
\def\nd{\end{eqnarray}}
\def\nn{\nonumber}

\def\ie{{\it i.e.}}

\def\nott#1{\setbox0=\hbox{$#1$}                
   \dimen0=\wd0                                 
   \setbox1=\hbox{/} \dimen1=\wd1               
   \ifdim\dimen0>\dimen1                        
      \rlap{\hbox to \dimen0{\hfil/\hfil}}      
      #1                                        
   \else                                        
      \rlap{\hbox to \dimen1{\hfil$#1$\hfil}}   
      /                                         
   \fi}                                         %

\newcommand{\beq}{\begin{eqnarray}}
\newcommand{\eeq}{\end{eqnarray}}

\newcommand{\p}{{\partial}}

\newcommand{\s}{{\sigma}}
\newcommand{\vev}[1]{{\langle #1 \rangle}}

\newcommand{\ord}[1]{{{\cal O}(#1)}}
\newcommand{\gappeq}{\mathrel{\rlap {\raise.5ex\hbox{$>$}}
{\lower.5ex\hbox{$\sim$}}}}
\newcommand{\lappeq}{\mathrel{\rlap{\raise.5ex\hbox{$<$}}
{\lower.5ex\hbox{$\sim$}}}}
\newcommand{\myref}[1]{(\ref{#1})}

\newcommand{\ben}{\begin{enumerate}}
\newcommand{\een}{\end{enumerate}}

\newcommand{\sbar}{{\bar \s}}

\newcommand{\psib}{{\bar \psi}}

\newcommand{\bit}{\begin{itemize}}
\newcommand{\eit}{\end{itemize}}

\newcommand{\Ncal}{{\cal N}}

\newcommand{\Ocal}{{\cal O}}

\newcommand{\sighat}{{\hat \s}}

\newcommand{\sbarhat}{{\hat \sbar}}

\newcommand{\mykill}[1]{}
\def\hf{\frac 12}

\def\[{\left [}
\def\]{\right ]}
\def\({\left (}
\def\){\right )}

\begin{document}

\begin{titlepage}
\renewcommand{\thepage}{}

\title{Lattice four-dimensional \NfourSYM\ is practical}

\author{Joshua W.~Elliott$^a$, 
Joel Giedt$^b$
and Guy D.~Moore$^a$}
\affiliation{$\,^a$
    Physics Department,
    McGill University,
    3600 rue University,
    Montr\'{e}al, QC H3A 2T8, Canada\\
    $\,^b$
    Department of Physics, Applied Physics, and Astronomy,
    Rensselaer Polytechnic Institute,
    110 Eighth Street, Troy, New York 12180-3590 USA
}

\date{30 May 2008}

\begin{abstract}
\noindent
We show that nonpertubative lattice studies of four-dimensional
\NfourSuYM\ are within reach.  We use Ginsparg-Wilson fermions
to avoid gluino masses and an exact implementation of the 
(chiral) $R$-symmetry, which greatly
limits the number of counterterms that must be fine-tuned. 
Only bosonic operators require fine tuning, so all tunings can be done
``offline'' by a
Ferrenberg-Swendsen type reweighting.  We show what measurables can be
used to perform the tuning.

\end{abstract}

\maketitle

\end{titlepage}

\renewcommand{\thepage}{\arabic{page}}

%
It is often assumed that the study of supersymmetry (SUSY) on the lattice
will be difficult in four dimensions because multiple relevant/marginal operators
that violate SUSY are allowed by lattice symmetries.
A notable exception is pure $\Ncal{=}1$ super-Yang-Mills (SYM). In that
case, lattice chiral symmetry in the form of Ginsparg-Wilson (GW) fermions
prevents additive renormalization of the gluino mass in the continuum limit,
and hence the only SUSY-violating operator is forbidden in that limit
by setting the bare mass to zero \cite{N1people}
\NfourSYM\
has scalars and Yukawa couplings, and so various masses and couplings
will receive divergent corrections in the continuum limit and must be
(nonperturbatively) fine-tuned away. Hence lattice \Nfour\
SYM has always seemed impractical by this fine-tuning approach.\footnote{Other
approaches recently suggested include those of Refs.~\cite{Catterall:2005fd,Kaplan:2005ta},
involving ``orbifold'' or ``twisted SUSY'' lattices.}

We show here that this is not the case. Using GW fermions 
the four gluinos can be kept massless and the $SU(4)_R$
symmetry in the lattice theory can be preserved, 
which greatly restricts the form of the
renormalizations. This leaves a scalar mass, two quartic couplings, and
a Yukawa coupling to tune. The Yukawa coupling can be tuned by
rescaling the scalar kinetic term (see below); %
so all tunings can be done by
adjusting bosonic terms in the action. This allows the tunings to be
done by the ``Ferrenberg-Swendsen method'' \cite{Falcioni:1982cz,FerrenbergSwendsen},
exploring a wide swath of coupling constant space ``offline'' from the results
of a single Monte-Carlo simulation. The parameter range available with
good statistics can be enlarged using multicanonical techniques
\cite{Baumann:1986iq,kajantie_electroweak}. Thus we arrive
at the encouraging result that all fine-tuning can be performed through an
``offline'' analysis, \ie, new simulations and fermion matrix inversions
are not required.


%
The continuum field content is $SU(N_c)$ YM theory
with an $SU(4)_R$ internal symmetry; there are
four Majorana fermions whose left handed components transform in the
fundamental ${\mathbf 4}$ representation
of $SU(4)_R$ and 6 real scalars in the antisymmetric tensor
representation. The six real scalars 
will be expressed with a single index $\phi_m$, $m{=}1{\dots}6$, or
composed into $SU(4)_R$
Weyl matrices:  $\phi_{ij}{=}\phi_m \sighat_{m,ij}$ and
$\phi^{ij} {=} \phi_m \sbarhat^{ij}_m$, where 
$\sighat$'s are just $SU(4)_R$ Clebsch-Gordon coefficients
involved in ${\bf 4^*} {\ni}\, {\bf 6} {\otimes} {\bf 4}$. 
The continuum action is
\bg
&&\!\!\!\!\!\!\!S = \frac{1}{g^2}\tr \big\{\hf F^2+|D_\mu\phi_m|^2
+\bar{\psi}_i \nott{D}\psi_i\\
&&\;\;\;\;+\sqrt{2}\bar{\psi}_i\left(\phi^{ij} P_L{-}(\phi^{ij})^*P_R\right)\psi_j
+[\phi_m,\phi_n][\phi_m,\phi_n]\big\}.\nn
\nd

The $SU(4) {\simeq} SO(6)$ preserving bosonic lattice action is a trivial
transcription from the continuum. Of course we must allow for generic
coefficients and non-SUSY terms, so that the SUSY-restoring
counterterms can be tuned. In our case these are entirely scalar terms.

It has been argued that 
naive lattice Yukawa terms lead to inconsistencies in either the 
chiral or Majorana projections (depending on how the Yukawas are 
transcribed to the lattice)~\cite{Fujikawa2002},
so the fermionic implementation is more subtle.
Following L\"uscher \cite{Luscher:1998pqa} and Kikukawa and Suzuki
\cite{Kikukawa:2004dd}, we introduce a set of auxiliary fermionic fields
$\Psi$, with lattice fermionic action
\bg
&&\hspace{-.3in}S_{\text{F}}
=a^4\tr\big\{ \bar\psi_i D_{\mbox{\tiny GW}} \psi_i - a^{-1}\bar\Psi_i\Psi_i\\
&&\hspace{.2in}+ y \sqrt{2}
(\bar\psi{+}\bar\Psi)_i\left(\phi^{ij} P_L-(\phi^{ij})^*P_R\right)
(\psi{+}\Psi)_j\big\}.\nn
\label{eqn:auxaction}
\nd
This action possesses an exact $SU(4)_R$ symmetry, with the scalars
transforming as in the continuum and the fermions transforming according
to
\bg
\hspace{-.2in}\begin{array}{l}
\delta\psi/i\eps=(T\hat{P}_{\!L}{-}T^*\hat{P}_{\!R})\psi \,, \\
\delta\Psi/i\eps=(T{+}T^*)\gamma_5D_{\mbox{\tiny GW}}\psi 
+ (TP_{\!L}{-}T^*P_{\!R})\Psi,\\ 
\delta\bar\psi/i\eps=\bar\psi\, (T^*P_{\!L}{-}TP_{\!R})
+ \bar\Psi (T{+}T^*)\gamma_5 \,, \\
\delta\bar\Psi/i\eps=-\bar\Psi\, (TP_{\!L}{-}T^*P_{\!R}) \,.
\end{array}
\nd
Here $\hat{P}_{L/R}\equiv\hf(1{\pm}\gfivehat)
=\hf(1{\pm}\gfive(1{-}2D_{\mbox{\tiny GW}}))$
are the lattice modified chiral projection operators, $T$ is the 
generator of $SU(4)_R$ in the fundamental ($\mathbf 4$)
and we have suppressed the $SU(4)_R$ indices.
Hence $(\psi{+}\Psi)$ and $(\bar\psi{+}\bar\Psi)$ transform like the
continuum $\psi,$ $\bar\psi$ fields.

This auxiliary field method preserves the $R$-symmetry exactly and keeps
the Yukawa terms ultralocal. It is also consistent with the Majorana
decomposition, so the fermionic determinant is an exact square; taking
its square root to implement the Majorana nature of the fermions retains
locality. The cost is the introduction of an extra fermionic excitation
$\Psi$, which is however nondynamical with $\ord{a^{-1}}$ mass, so it
decouples from the theory in the continuum limit.
Note however that in building lattice operators which contain fermions,
we must always use the combination $\psi{+}\Psi$, since this is the
combination with correct $R$-symmetry transformations.

%
%
We are interested in tuning the lattice action such that the effective
infrared description is \NfourSYM. Generically there is a nontrivial
matching between lattice and effective IR theories and all
relevant or marginal terms consistent with lattice symmetries
will appear in the infrared, except at special points in
bare parameter space. We can arrive at the desired special
point (\ie, \NfourSYM) by introducing the SUSY-violating operators
into the bare action and fine-tuning counterterms. 
These counterterms fall into three categories:
a scalar mass term, a Yukawa term, and two or four scalar quartic terms,
depending on the number of colors for the gauge group,
restricted here to $SU(\nc)$.
In particular, if $\nc\leq 3$ then $SU(4)_R$ symmetry allows only two
unique quartic terms:
$(\tr \phi_m \phi_m)^2$ and $(\tr \phi_m \phi_n)^2$ (trace over gauge
indices, $R$-indices explicit).
For $\nc>3$ the quartic invariants
$\tr \phi_m \phi_m \phi_n \phi_n$ and $\tr \phi_m \phi_n \phi_m \phi_n$
are also independent (for $\nc\leq 3$ they are linearly dependent on the
first two).

Rescaling the Yukawa term,
schematically $y \psib \phi \psi$,
can be accomplished through a rescaling of the scalar kinetic term.
Consider replacing $y\psib \phi \psi + |D\phi|^2$ with
$y\psib\phi\psi + Z |D\phi|^2$.  In terms of the canonically normalized
scalar field this is $(y/\sqrt{Z}) \psib\phi\psi + |D\phi|^2$.  Therefore
we include a counterterm for the kinetic term rather than for the Yukawa.
(Note that the rescaling of the scalar kinetic term also rescales the
scalar potential; but there are distinct counterterms to undo this
rescaling separately).


Suppose we perform a Monte Carlo simulation at one value $m_1$ of the scalar mass $m$,
so that the configurations sample the distribution determined
by the action $S= S_{m=0} {+} \hf \int\! m_1^2 \phi^2$.
Following the ``Ferrenberg-Swendsen reweighting'' method 
\cite{Falcioni:1982cz,FerrenbergSwendsen}
one can use the following
``reweighting identity'' to compute the expectation value of an operator
$O$ for the distribution with a mass $m_2$:
\beq
\vev{ \Ocal } = \mbox{$\sum_c$} \Ocal_c  e^{ -\hf (m_2^2-m_1^2) \int\! \phi^2_c } 
\Big/ \mbox{$\sum_c$} e^{ -\hf (m_2^2-m_1^2) \int\! \phi^2_{c}}.
\label{ilte}
\eeq

There is a limited regime of utility to this technique, due to
the so-called ``overlap problem.''  For instance,
if the exponential in \myref{ilte} is large where the
simulated distribution has little weight, a finite sampling
will have large errors. The mismatch of the distributions
gets worse as the number of lattice sites increases, because
the exponent is extensive.

A way to ameliorate the overlap problem,
which has been found to work in other contexts, is
``multicanonical reweighting'' \cite{Baumann:1986iq}.
One replaces $S$ with  $S+W[O_1,O_2,\ldots]$, 
where $W[O_1,O_2,\ldots]$ is a
carefully chosen function of some small set of observables (in our case
$W$ will be a function of $\int\! \phi^2$, the distinct $\int\! \phi^4$'s,
and $\int\! (D \phi)^2$).
The expectation value of an observable in the distribution corresponding
to $S$ is:
\bg
\langle {\cal O} \rangle = \mbox{$\sum_c$}{\cal O}_c \; e^{W[O^c_1,...]}
\Big/
                               \mbox{$\sum_c$} e^{W[O^c_1,...]}
\nd
$W[..]$ produces a weighted average
over a continuum of canonical ensembles, 
some of which will have a good overlap with the distribution that one is
reweighting to. The challenge is to design a $W$ such that
sampling is flattened over the range of observables one is interested in.


Two approaches to engineering a good function $W$ exist:
(1) a bootstrap method that iterates
between Monte Carlo simulation and adjusting $W$, and (2)
optimizing $W$ w.r.t.~its parameters, in a small volume,
and then using step-scaling to extrapolate to a good estimate
for $W$ in larger volumes. For instance one can start with $4^4$ and
$6^4$ volumes, where statistics accumulate rapidly and unreweighted
simulations still cover broad parameter ranges.


%
In our case the reweighting function $W$ will depend on the four bosonic
contributions to the action, $\int\!\phi^2$, $\int\!(D\phi)^2$,
$\int\!\phi_1^4$ and $\int\!\phi^4_2$
(the two quadratic and two independent quartic operators you can form from
the scalars, integrated over space).
By sampling with the weight
\bg
\rho={\cal D}[A,\phi]\det[D] 
\;e^{-S[A,\,\phi\,;\,\,m^2_0,\dots]}
\;e^{-W[\int\!\phi^2,\dots]}\,,
\nd
and you can reproduce the ensemble at some particular set of values for
$m^2$, $Z_\phi$, $\lambda$ and $\lambda'$, via
\bg
Z = \mbox{$\sum_c$} \,e^{+W[\dots]}
 \;e^{        - [
                (m^2 {-}m_0^2) \phi^2_c
              + (Z_\phi{-}Z^0_\phi)(D\phi)^2_c
              + \dots
              ]}\,,
\nd
with $W$ chosen so that the sample has a
reasonable number of configurations for all values of $\int\!\phi^2$,
$\int\!(D\phi)^2$, $\int\!\phi^4$ and $\int\!(\phi^2)^2$ within some 
interesting range.

Now, define the gauge invariant effective potential in finite
volume as follows:
\bg
\!e^{-\Omega \;V[A^2]} \!= \!\mbox{$\sum_c$}\,e^{+W[\ldots]}
 \;e^{- [ (m^2 {-}m_0^2) \phi^2_c\dots]}
 {\times} \delta(A^2 {-} \frac{\phi^2_c}{\Omega})
\nd
with $\Omega$ the 4-volume (so that $A^2$ represents the mean value of
the squared scalar field).
It is easy to check that varying $m^2$ changes $V[A^2]$ by adding a
linear component. Therefore, measuring $V[A^2]$ immediately
determines how $\langle \phi^2 \rangle$ and $F{=}\ln Z$ vary as a function
of $m^2$.  One easily generalizes to the effective potential as a
function of all four bosonic operators, which gives a quick way to
explore the effect of varying parameters.

It remains to specify how to tune the parameters.
Consider \NfourSYM\ with an added $a^2(\Tr \phi_m \phi_m)^3$ term in the
potential, but deformed by mass and quartic interactions.  The SUSY
point is a second order phase transition point; for negative quartic
deformation there is a first order transition as $m^2$ is varied (the
system jumps from a massive state about the origin to a massive state
with a large VEV because the quartic term bends down).  For positive
quartic deformation there is no phase transition.  The optimal value of
$m^2$ (fixing other parameters) is the point of maximal susceptibility
$\langle (\int_x \phi^2)^2\rangle -(\langle \int_x \phi^2 \rangle)^2$.
Determining this point in finite volume leads to $\ord{1/L^2}$ errors in
the determined value of $m^2$, which can be improved by scaling over multiple
volumes.  Finding the flat quartic term which gives second-order
behavior should also be possible; it has been successfully achieved in
the context of the electroweak phase transition \cite{Kajantie}.
Therefore it should be possible to use the phase diagram to tune at
least two parameters.  Note that we needed to add a lattice-size
$\phi^6$ term; this is harmless but it raises the issue that
the flat direction actually means that unbroken \NfourSYM\ is
not well behaved in finite volume; the moduli are not fixed and the
partition function diverges because of the integral over the infinite
moduli space. Therefore it will always be necessary to break SUSY
somehow. We advocate doing so via twisted boundary conditions; for
instance, instead of periodic boundary conditions we can add 
a rotation by angle $\Theta$ to all fermionic fields in one direction.
The choice
$\Theta{=}\pi$ is the maximal global breaking of SUSY and corresponds to
treating the thermal ensemble; intermediate values of $\Theta$
break SUSY by smaller amounts. This lifts the moduli degeneracy without
any {\em local} SUSY breaking; the effects of $\Theta$ are only visible
in correlations at the scale of the lattice size, which is anyway
contaminated by being in finite volume.

For the case $\nc>3$ the effective potential should show multiple flat
directions in the space of quartic operators; only one quartic direction
(some linear combination of the input quartics, due to mixing) should
rise steeply.  Therefore we expect it should be possible to tune the
``extra'' quartic operators in the case $\nc>3$, leaving only one
quartic and the Yukawa coupling/wave function to tune.

%
If SUSY is exact then the ($R$-symmetry {\bf 4})
supercurrent $S_{\mu,i}$
is conserved, so $\langle \partial^\mu S_{\mu,i}(x) \Ocal(y) \rangle$
vanishes at $x{\neq} y$ for all local operators $\Ocal$. We can use this
property to measure whether we are at the SUSY point in parameter space,
and therefore to tune parameters to find the SUSY point. The technique
has been pioneered in ${\cal N}{=}1$ SUSY with Wilson fermions by the
DESY-M\"unster group \cite{Farchioni:2001wx}; here we discuss the extension
to \NfourSYM.

The supercurrent $S_{\mu,i}$ is a linear combination of 
three dimension-7/2 operators. It is easy to find lattice operators
which reproduce these continuum operators, at tree level and with
contamination from higher dimension operators. The choices are not
unique and at the nonperturbative level each lattice operator will mix
with all continuum operators in the same symmetry channel.
Different choices of lattice
operator will reproduce the continuum operator with different
normalization, mixings, and $\ord{a}$ suppressed higher dimension
contamination.
Hence we express the operators $\Ocal_{\mu,i}$
in a continuum language, and leave the particulars of lattice
transcription (which amounts to various ``improvements'' w.r.t.~$\ord{a}$
discretization errors) for detailed studies. Our intention here
is to lay out the methodology.

In their analysis of the $\Ncal{=}1$ SYM case, the DESY-M\"unster group
found two dimension-7/2 operators, the supercurrent $S_{\mu}$ and
another fermionic current $T_\mu$. These mix in the lattice-continuum
matching and so one must write down two lattice operators with
undetermined coefficients in order to find something which corresponds
purely to $S_\mu$ (plus $\ord{a}$ dimension-9/2 contamination).
In the present \Nfour\ case there are 5 dimension-7/2 operators which we
will name $\Ocal_{\mu,i}^{1\ldots 5}$, and the renormalized \Nfour\
supercurrent will, in all generality, take the form:
\beq&&
\begin{array}{l}\vspace{.05in}
\hspace{-.2in}S_{\mu,i}^{\text{ren.}}=
 \big\{ Z_1 \frac{1}{2}F_A{\cdot}\sigma\;\delta_{ij}
+ Z_2 \sqrt{2}\nott{D}\big(\phi_{ij}P_{\!L}{+}\phi^{ij} P_{\!R}\big)_A\\
\vspace{.05in}\hspace{.3in}
- Z_3 f^{ABC}\big(\phi^B_{ik}\phi^{kj}_C P_{\!R}+\phi^{ik}_B\phi^C_{kj}P_{\!L}\big)\big\}\gamma_\mu\psi_{\!jA} \\
\hspace{-.2in}+ \big\{ \!Z_4 \gamma_\nu F^A_{\mu\nu}\delta_{ij}
- Z_5 \sqrt{2}D_\mu \big(\phi^{ij}P_{\!L}+\phi_{ij}P_{\!R}\big)_{\!\!A}\!\big\} 
\psi_{\!jA} + \ord{a} \end{array}\nn\\&&
\hspace{.2in} \equiv  Z_n \Ocal^n_{\mu,i} + \ord{a},
\eeq
where the terms on the righthand side are bare (lattice) operators.
Note that:  (1) at tree level the supercurrent corresponds
to $Z_1{=}Z_2{=}Z_3{=}1$ and $Z_4{=}Z_5{=}0$; (2) the renormalization constants
$Z_{n}$ are universal w.r.t.~the index $i$ due
to the $SU(4)_R$ symmetry preserved by the lattice.

We can tune to the SUSY point by varying parameters to force
correlation functions of this lattice-implemented 
$\p_\mu S_{\mu,i}$ to vanish up to $\ord{a}$ corrections.
Specifically, to tune two parameters we need to choose 6 operators
$\Ocal^n_{\mu,i}$ in the same symmetry channel as $S_{\mu,i}$ (otherwise the
correlation function vanishes automatically). The natural choice is
$\Ocal^{1,\ldots 5}_{\mu,i}$ plus one dimension-9/2 operator
$\Ocal^6_{\mu,i}$. One then measures the matrix of correlation functions
\st
M^{mn}(t) \equiv \int d^3 \vec{x}
\langle \Ocal^{m\dagger}_{0,i}(t,\vec{x}) \Ocal^n_{0,i}(0,0) \rangle
\stp
whose $t$ derivative is the correlation function between $\p_\mu
\Ocal^m_{\mu,i}$ and $\Ocal^n(0,i)$ at vanishing spatial momentum. Since the
operators involved are dimension-7/2 we generically expect the elements
of $M^{mn}(t)$ to decay as $t^{-7}$. At the SUSY point and
for the right choices of $Z_m$, $Z_m M^{mn}$ decays as $at^{-8}$ for all
$n$. We can fix the undetermined ratios $Z_{2\ldots 5}/Z_1$ by
enforcing that this holds for $n=1\ldots 4$. Forcing that it hold for
$n{=}5,6$ gives two conditions which can be used to check whether we are
at the SUSY point--tuning to the SUSY point is tuning for
$Z_m M^{m5}{\sim} at^{-8}$ and $Z_m M^{m6}{\sim} at^{-8}$. Actually since
one of the operators is dimension-9/2 we must force one linear
combination $Z_m M^{mn} c_n$ to vanish as $a^2t^{-9}$.

We do not see an obstacle to using this procedure to tune more
parameters, if it proves too difficult to tune some of the vanishing
quartic couplings via the potential method. Therefore in principle the
tuning to the SUSY point can be done by any mixture of the Ward identity
method and the effective potential method.

%
We have seen that for $\nc{=}2,3$ colors, there are four fine-tunings
in the action. For $\nc{>}3$ colors there are six. In addition, one
must fix the four relative renormalization constants in the
supercurrent. All but one of the scalar potential counterterms can
be fixed by matching the effective potential, as
determined by the multicanonical simulation, to the target
theory scalar potential $\tr [\phi_m, \phi_n] [\phi_m, \phi_n]$.
The overall strength of this term cannot be determined from the
effective potential, because it will be expressed in terms of the
bare operators in our approach.

This leaves just six fine-tunings for all number of colors $\nc$:
one fine-tuning of the bare kinetic coefficient for the scalar,
one overall scalar potential coefficient, and
the four relative supercurrent coefficients. Thus a total
of six Ward identities must be measured well enough to distinguish
their simultaneous minimum w.r.t.~$Z_\phi, Z_1/Z_2, \ldots, Z_4/Z_5$.

\medskip

\centerline{\bf Discussion}

\medskip
By preserving the $SU(4)_R$ symmetry of the target theory, the
number of counterterms that must be fine-tuned is greatly reduced.
This can be done by implementing GW 
fermions, with the chirality of
the Yukawa couplings implemented with auxiliary fermions, extending the
method of \cite{Kikukawa:2004dd}.
Because counterterm fine-tuning can be isolated to the purely
bosonic sector, it can all be done off-line, \ie, without the
expense of fermion matrix inversions (the bottleneck for all
dynamical fermion simulations). This is a great advantage, because
a very large number of points in the bare action parameter space
will have to be scanned in order to find the \NfourSYM\ point.
Finally, we have explained how the overlap problem can be alleviated
by taking a multicanonical approach, flattening the distributions that
will be scanned over.

The main limitation to this method 
is that, since \NfourSYM\ is conformal, the
continuum limit is not a weak coupling limit. Our proposal should work
at weak coupling, where one knows that the infrared description will be
in terms of the same degrees of freedom as one puts on the lattice. But
there is no guarantee that one can find lattice parameters which
correspond to strongly coupled continuum theories.

The principal challenge is that the method requires GW fermions,
which are numerically expensive--especially in a theory such as this
one, with massless particles and the corresponding critical slowing
down. It will be a challenge to generate enough configurations
to measure quantities with sufficient accuracy to determine the
SUSY point. Fermions are necessarily involved in the correlation
functions of the supercurrent, so storage of propagators during the simulations will
be essential to performing the fine-tuning w.r.t.~Ward identities.
The storage and computing resources that will be required
will be substantial, but we believe that 
the exploratory studies that need to be
done can be performed in the near term. 
For instance, one of the authors (JG) has access to
the Computational Center for Nanotechnology
Innovation (CCNI) at Rensselaer. Indeed, it is
currently being used by Giedt and collaborators
for $\Ncal{=}1$ SYM 
simulations 
at a sustained actual compute rate of 1 Tflop/s, 
precisely the sort of resource that would be
required to perform exploratory studies of this proposal.
%
Obviously early stages
of such work will be very much technical studies of the lattice
theory. Continuum results will take much longer. Nevertheless,
the beginnings of first principles nonperturbative study of
\Nfour\ SYM are not so far off, we believe, if the current proposal
is pursued with some dedication and adequate resources. We
hope to report on further progress in that direction in the
near future.

\smallskip

\centerline{\bf Acknowledgements}
\noindent This work was supported in part by
the Natural Sciences and Engineering Research Council of Canada.

\bibliographystyle{apsrev}
\bibliography{prlonebib}

\begin{thebibliography}{20}
\expandafter\ifx\csname natexlab\endcsname\relax\def\natexlab#1{#1}\fi
\expandafter\ifx\csname bibnamefont\endcsname\relax
  \def\bibnamefont#1{#1}\fi
\expandafter\ifx\csname bibfnamefont\endcsname\relax
  \def\bibfnamefont#1{#1}\fi
\expandafter\ifx\csname citenamefont\endcsname\relax
  \def\citenamefont#1{#1}\fi
\expandafter\ifx\csname url\endcsname\relax
  \def\url#1{\texttt{#1}}\fi
\expandafter\ifx\csname urlprefix\endcsname\relax\def\urlprefix{URL }\fi
\providecommand{\bibinfo}[2]{#2}
\providecommand{\eprint}[2][]{\url{#2}}

\bibitem[{\citenamefont{N1people}(2000)}]{N1people}
\bibinfo{author}{\bibfnamefont{G.}~\bibnamefont{Curci}} \bibnamefont{and}
  \bibinfo{author}{\bibfnamefont{G.}~\bibnamefont{Veneziano}},
  \bibinfo{journal}{Nucl. Phys.} \textbf{\bibinfo{volume}{B292}},
  \bibinfo{pages}{555} (\bibinfo{year}{1987}).
\bibinfo{author}{\bibfnamefont{N.}~\bibnamefont{Maru}} \bibnamefont{and}
  \bibinfo{author}{\bibfnamefont{J.}~\bibnamefont{Nishimura}},
  \bibinfo{journal}{Int. J. Mod. Phys.} \textbf{\bibinfo{volume}{A13}},
  \bibinfo{pages}{2841} (\bibinfo{year}{1998}).
\bibinfo{author}{\bibfnamefont{H.}~\bibnamefont{Neuberger}},
  \bibinfo{journal}{Phys. Rev.} \textbf{\bibinfo{volume}{D57}},
  \bibinfo{pages}{5417} (\bibinfo{year}{1998}).
\bibinfo{author}{\bibfnamefont{J.}~\bibnamefont{Nishimura}},
  \bibinfo{journal}{Phys. Lett.} \textbf{\bibinfo{volume}{B406}},
  \bibinfo{pages}{215} (\bibinfo{year}{1997}).
\bibinfo{author}{\bibfnamefont{D.~B.} \bibnamefont{Kaplan}} \bibnamefont{and}
  \bibinfo{author}{\bibfnamefont{M.}~\bibnamefont{Schmaltz}},
  \bibinfo{journal}{Chin. J. Phys.} \textbf{\bibinfo{volume}{38}},
  \bibinfo{pages}{543} (\bibinfo{year}{2000}).
\bibinfo{author}{\bibfnamefont{I.}~\bibnamefont{Campos}} \bibnamefont{et~al.}
  (\bibinfo{collaboration}{DESY-Munster}), \bibinfo{journal}{Eur. Phys. J.}
  \textbf{\bibinfo{volume}{C11}}, \bibinfo{pages}{507} (\bibinfo{year}{1999}).
\bibinfo{author}{\bibfnamefont{G.~T.} \bibnamefont{Fleming}},
  \bibinfo{author}{\bibfnamefont{J.~B.} \bibnamefont{Kogut}}, \bibnamefont{and}
  \bibinfo{author}{\bibfnamefont{P.~M.} \bibnamefont{Vranas}},
  \bibinfo{journal}{Phys. Rev.} \textbf{\bibinfo{volume}{D64}},
  \bibinfo{pages}{034510} (\bibinfo{year}{2001}).
\bibinfo{author}{\bibfnamefont{R.}~\bibnamefont{Kirchner}},
  \bibinfo{author}{\bibfnamefont{S.}~\bibnamefont{Luckmann}},
  \bibinfo{author}{\bibfnamefont{I.}~\bibnamefont{Montvay}},
  \bibinfo{author}{\bibfnamefont{K.}~\bibnamefont{Spanderen}},
  \bibnamefont{and}
  \bibinfo{author}{\bibfnamefont{J.}~\bibnamefont{Westphalen}}
  (\bibinfo{collaboration}{DESY-Munster}), \bibinfo{journal}{Nucl. Phys. Proc.
  Suppl.} \textbf{\bibinfo{volume}{73}}, \bibinfo{pages}{828}
  (\bibinfo{year}{1999}).
\bibinfo{author}{\bibfnamefont{A.}~\bibnamefont{Donini}},
  \bibinfo{author}{\bibfnamefont{M.}~\bibnamefont{Guagnelli}},
  \bibinfo{author}{\bibfnamefont{P.}~\bibnamefont{Hernandez}},
  \bibnamefont{and} \bibinfo{author}{\bibfnamefont{A.}~\bibnamefont{Vladikas}},
  \bibinfo{journal}{Nucl. Phys.} \textbf{\bibinfo{volume}{B523}},
  \bibinfo{pages}{529} (\bibinfo{year}{1998}).

\bibitem[{\citenamefont{Falcioni et~al.}(1982)\citenamefont{Falcioni, Marinari,
  Paciello, Parisi, and Taglienti}}]{Falcioni:1982cz}
\bibinfo{author}{\bibfnamefont{M.}~\bibnamefont{Falcioni}},
  \bibinfo{author}{\bibfnamefont{E.}~\bibnamefont{Marinari}},
  \bibinfo{author}{\bibfnamefont{M.~L.} \bibnamefont{Paciello}},
  \bibinfo{author}{\bibfnamefont{G.}~\bibnamefont{Parisi}}, \bibnamefont{and}
  \bibinfo{author}{\bibfnamefont{B.}~\bibnamefont{Taglienti}},
  \bibinfo{journal}{Phys. Lett.} \textbf{\bibinfo{volume}{B108}},
  \bibinfo{pages}{331} (\bibinfo{year}{1982}).

\bibitem[{\citenamefont{Ferrenberg and Swendsen}(1988)}]{FerrenbergSwendsen}
\bibinfo{author}{\bibfnamefont{A.~M.} \bibnamefont{Ferrenberg}}
  \bibnamefont{and} \bibinfo{author}{\bibfnamefont{R.~H.}
  \bibnamefont{Swendsen}}, \bibinfo{journal}{Phys. Rev. Lett.}
  \textbf{\bibinfo{volume}{61}}, \bibinfo{pages}{2635} (\bibinfo{year}{1988}).

\bibitem[{\citenamefont{Baumann}(1987)}]{Baumann:1986iq}
\bibinfo{author}{\bibfnamefont{B.}~\bibnamefont{Baumann}},
  \bibinfo{journal}{Nucl. Phys.} \textbf{\bibinfo{volume}{B285}},
  \bibinfo{pages}{391} (\bibinfo{year}{1987}).

\bibitem[{\citenamefont{Kajantie et~al.}(1996)\citenamefont{Kajantie, Laine,
  Rummukainen, and Shaposhnikov}}]{kajantie_electroweak}
\bibinfo{author}{\bibfnamefont{K.}~\bibnamefont{Kajantie}},
  \bibinfo{author}{\bibfnamefont{M.}~\bibnamefont{Laine}},
  \bibinfo{author}{\bibfnamefont{K.}~\bibnamefont{Rummukainen}},
  \bibnamefont{and} \bibinfo{author}{\bibfnamefont{M.~E.}
  \bibnamefont{Shaposhnikov}}, \bibinfo{journal}{Nucl. Phys.}
  \textbf{\bibinfo{volume}{B466}}, \bibinfo{pages}{189} (\bibinfo{year}{1996}).

\bibitem[{\citenamefont{Fujikawa et~al.}(2002)\citenamefont{Fujikawa,
  Ishibashi, and Suzuki}}]{Fujikawa2002}
\bibinfo{author}{\bibfnamefont{K.}~\bibnamefont{Fujikawa}},
  \bibinfo{author}{\bibfnamefont{M.}~\bibnamefont{Ishibashi}},
  \bibnamefont{and} \bibinfo{author}{\bibfnamefont{H.}~\bibnamefont{Suzuki}},
  \bibinfo{journal}{Phys. Lett.} \textbf{\bibinfo{volume}{B538}},
  \bibinfo{pages}{197} (\bibinfo{year}{2002}).

\bibitem[{\citenamefont{Luscher}(1998)}]{Luscher:1998pqa}
\bibinfo{author}{\bibfnamefont{M.}~\bibnamefont{Luscher}},
  \bibinfo{journal}{Phys. Lett.} \textbf{\bibinfo{volume}{B428}},
  \bibinfo{pages}{342} (\bibinfo{year}{1998}).

\bibitem[{\citenamefont{Kikukawa and Suzuki}(2005)}]{Kikukawa:2004dd}
\bibinfo{author}{\bibfnamefont{Y.}~\bibnamefont{Kikukawa}} \bibnamefont{and}
  \bibinfo{author}{\bibfnamefont{H.}~\bibnamefont{Suzuki}},
  \bibinfo{journal}{JHEP} \textbf{\bibinfo{volume}{02}}, \bibinfo{pages}{012}
  (\bibinfo{year}{2005}).

\bibitem[{\citenamefont{Rummukainen et~al.}(1998)}]{Kajantie}
\bibinfo{author}{\bibfnamefont{K.}~\bibnamefont{Rummukainen}}
  \bibnamefont{et~al.}, \bibinfo{journal}{Nucl. Phys.}
  \textbf{\bibinfo{volume}{B532}}, \bibinfo{pages}{283} (\bibinfo{year}{1998}),
  \eprint{hep-lat/9805013}.

\bibitem[{\citenamefont{Farchioni et~al.}(2002)}]{Farchioni:2001wx}
\bibinfo{author}{\bibfnamefont{F.}~\bibnamefont{Farchioni}}
  \bibnamefont{et~al.} (\bibinfo{collaboration}{DESY-Munster-Roma}),
  \bibinfo{journal}{Eur. Phys. J.} \textbf{\bibinfo{volume}{C23}},
  \bibinfo{pages}{719} (\bibinfo{year}{2002}).

\bibitem[{\citenamefont{Catterall}(2005)}]{Catterall:2005fd}
\bibinfo{author}{\bibfnamefont{S.}~\bibnamefont{Catterall}},
  \bibinfo{journal}{JHEP} \textbf{\bibinfo{volume}{06}}, \bibinfo{pages}{027}
  (\bibinfo{year}{2005}).

\bibitem[{\citenamefont{Kaplan and Unsal}(2005)}]{Kaplan:2005ta}
\bibinfo{author}{\bibfnamefont{D.~B.} \bibnamefont{Kaplan}} \bibnamefont{and}
  \bibinfo{author}{\bibfnamefont{M.}~\bibnamefont{Unsal}},
  \bibinfo{journal}{JHEP} \textbf{\bibinfo{volume}{09}}, \bibinfo{pages}{042}
  (\bibinfo{year}{2005}).

\end{thebibliography}

\end{document}